Machine Learning Prediction Models for Solid Electrolytes based on Lattice Dynamics Properties

Jiyeon Kim[1,2], Donggeon Lee[3,4], Dongwoo Lee[5], Xin Li[6], Yea-Lee Lee,[7] Sooran Kim[1,8*]

[1]Department of Physics Education, Kyungpook National University, Daegu 41566, South Korea

[2]The Center for High Energy Physics, Kyungpook National University, Daegu 41566, South Korea

[3]Department of Physics, Kyungpook National University, Daegu 41566, South Korea

[4]SKKU Advanced Institute of Nanotechnology (SAINT) and Department of Nano Engineering, Sungkyunkwan University, Suwon 16419, South Korea

[5]School of Mechanical Engineering, Sungkyunkwan University, Suwon 16419, South Korea

[6]John A. Paulson School of Engineering and Applied Sciences, Harvard University, Cambridge, Massachusetts 02138, USA

[7]Chemical Data-Driven Research Center, Korea Research Institute of Chemical Technology, Daejeon 34114, South Korea

[8]KNU LAMP Research Center, KNU Institute of Basic Sciences, Kyungpook National University, Daegu, 41566 South Korea

*Corresponding authors: sooran@knu.ac.kr

## ABSTRACT

Recently, machine-learning approaches have accelerated computational materials design and the search for advanced solid electrolytes. However, the predictors are currently limited to static structural parameters, which may not fully account for the dynamic nature of ionic transport. In this study, we meticulously curated features considering dynamic properties and developed machine-learning models to predict the ionic conductivity of solid electrolytes. We compiled 14 phonon-related descriptors from first-principles phonon calculations along with 16 descriptors related to structure and electronic properties. Our logistic regression classifiers exhibit an accuracy of 93 %, while the random forest regression model yields a root mean square error of 1.179 S/cm and $R^2$ of 0.710. Notably, phonon-related features are essential for estimating the ionic conductivity in both models. Furthermore, we applied our prediction model to screen 264 Li-containing materials and identified 11 promising candidates as potential superionic conductors.

## TOC GRAPHIC

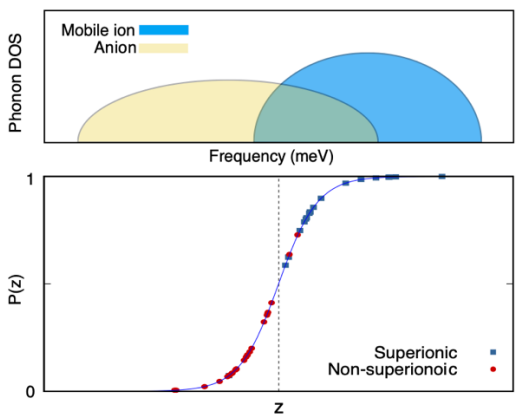

Concomitant to the sharp increase in global demand for electric vehicles, mobile electronics, and large energy storage, lithium-ion batteries (LIBs) have been intensively and extensively studied to improve their performance.[1–3] However, organic liquid electrolytes, commonly used in LIBs due to their high ionic mobility, pose potential safety risks.[4,5] Thermal instability of LIBs often arises from the breakage of the separator and electrochemical reactions within the electrolytes.

Inorganic solid-state electrolytes (SSE) have been investigated to mitigate the safety risks in the past decades.[6–11] Compared to liquid electrolytes, SSEs are advantageous in electrochemical and thermal stability,[10] as well as good cycle performance.[9,11] It is also possible to operate under high voltage and achieve large energy density by use of the metallic lithium anode and the high voltage cathode.[9,12] Examples include LISICON type (lithium superionic conductor), NASICON type (sodium superionic conductor), garnet, perovskites, and argyrodites materials such as $Li_{10}GeP_2S_{12}$,[13] $Li_{1.3}Al_{0.3}Ti_{1.7}(PO_4)_3$,[14] $Li_7La_3Zr_2O_{12}$,[15] $La_{0.5}Li_{0.5}TiO_3$,[16] and $Li_6PS_5Br$.[17] However, SSEs exhibit relatively low ionic conductivity compared to organic liquid electrolytes, therefore many studies have searched for SSEs with high ionic conductivity and electrochemical stability.[6,18,19]

Previous studies have suggested various parameters to control the ionic conductivity of solid electrolytes or superionic conductors. The static structure parameters[6] such as bottleneck size,[20,21] volume,[22,23] and anion sublattice[23] have been reported to be related to ionic transport and activation energy. Not only static properties but also dynamic properties have been suggested as important features for ion mobility.[6,24–31] For example, low energy optical (LEO) phonons induce a decrease in the activation barrier of ionic transport.[25] Soft lattice modes related to the octahedral rotation are connected to ionic diffusion in Ruddlesden-Popper phase $Ln_2NiO_{4+\delta}$ (Ln = La, Pr, Nd).[26]

Phonon instability and migration profiles of $Li_2O$ were discussed under both ambient and superionic phases.[27] Krauskopf et al. showed that the lattice softness reduces the activation energy for Na ion motion in $Na_3PS_{4-x}Se_x$.[28] The average phonon frequency of Li-ion was reported to correlate with the enthalpy of migration in LISICON,[29] and used as a descriptor for high-throughput screening.[32]

Theoretical approaches to ionic conductivity have traditionally relied on direct methods, such as *ab initio* molecular dynamics (MD),[33–35] despite being notably time-consuming. In recent years, the field of materials science has experienced a growing interest in the application of machine learning (ML) approaches,[36–54] aimed at predicting properties based on the accumulated data. Consequently, several studies have applied various ML techniques to design potential SSEs.[43–54] For example, Fujimura et al. performed the support vector regression to predict the ionic conductivity of LISICON.[54] Artificial neural network modeling was used to predict the Li diffusion barrier for $LiMXO_4$.[47] Sendek et al. developed a classification model using various structural types of Li-ion conductors as a training set and proposed 21 promising candidates for SSE.[48] Unsupervised learning and compositionally restricted attention based networks (CrabNets) were also applied to distinguish fast Li-ion conductors.[44,45]

In previous ML investigations,[45–52] common features based on the atomic structures and chemical properties of components were used due to their straightforward quantification and ease of data collection. However, features related to lattice dynamics or phonon properties have recently shown importance in correlation with ionic diffusion,[6,24–31] which have not been employed to predict the ion conductivity of the SSEs through ML approaches. Clarifying the mechanisms of ionic conductivities at fixed temperatures, such as room temperature, is challenging due to structural complexity and interactions. Therefore, it would be worth developing ML models with features that include both dynamic and structural properties of materials and investigating their contribution to ionic conductivity.

In this study, we have developed ML models to predict the ionic conductivity of SSEs by utilizing the features related to not only static properties such as structural, electronic, and chemical features but also lattice dynamics. Figure 1 illustrates the schematic workflow of this work. First, we collected experimental ionic conductivity at room temperature (RT) and the corresponding crystal structures of Li and Na-based SSE materials from previous literature as shown in Table S1. For feature derivation, we performed density functional theory (DFT) calculations and obtained thirty features related to the phonon, electronic, and structural properties. The logistic regression (LR) classifier and random forest (RF) regression were employed for classification and regression models, respectively. The contribution of the features to estimate ionic conductivity is quantified by the Gini importance (GI) in the RF model. Furthermore, based on the developed LR model, we screened Li-SSE candidates containing O or S anions using the Materials Project (MP) database[55] and the phonon database at Kyoto University.[56–58] Among them, 11 materials were identified as promising candidates for superionic conductors.

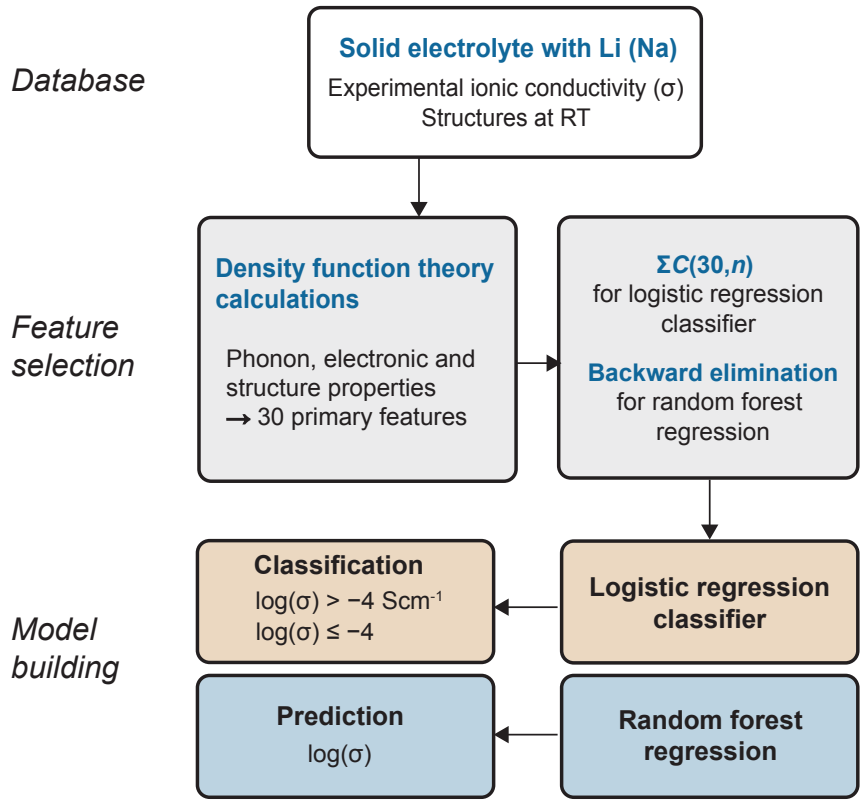

**Figure 1**. Schematic workflow of the ML model for ionic conductivity of Li and Na-based solid-state electrolytes.

To derive the features of electronic and structural properties, we conducted DFT calculations using the Vienna *ab initio* Simulation Package (VASP) code.[59,60] All calculations are performed using Perdew-Burke-Ernzerhof (PBE) functional of generalized gradient approximation (GGA)[61] and plane-wave energy cutoff of 650 eV. A *k*-point density of 5000/atom is used with respect to the corresponding system. The crystal structures were referenced from the MP[55] and Inorganic Crystal Structure Database (ICSD).[62] The atomic and lattice structures are fully relaxed until the Hellmann-Feynman force is less than 0.001 eV/Å while maintaining the initial symmetry of the experimental structures. The disorder structures that have partial occupancies were regenerated in ordered structure using Python Materials Genomics (pymatgen).[63] To obtain the phonon properties, we used the PHONOPY package, which applies a finite displacement approach and supercell calculations.[64] The force constants were calculated using the supercells with lattice parameters greater than 10 Å.

We employed LR and RF machine learning algorithms to classify and predict ionic conductivity. All models were developed and implemented through the application of the scikit-learn python library. We used the leave-one-out cross-validation (LOOCV) and evaluated models to suppress overfitting. The metric of accuracy and F1 score, and metrics of the root mean square error (RMSE) and coefficient of determination ($R^2$) were utilized for LR classification and RF regression models, respectively (See the Supporting information for the detailed information on accuracy and F1 score).

The LR model accepts any real value and generates a probability value between 0 and 1 through the logistic function $P(z) = \frac{1}{1+ e^{-z}}$. This model is applied in predicting ionic conductivity using the equation, so-called $z$ function, $z = w_0 + \sum_i w_i x_i$, where $w_i$ and $x_i$ denote the regression coefficient and the value of the feature, respectively. To find the best combination of features, we searched all combinations of 30 features, resulting in a total of $\Sigma \binom{30}{n} = 1{,}073{,}741{,}824$ models.

We implemented the RF model,[65] an ensemble of models using decision trees. By constructing a forest comprising numerous decision trees, the RF model generates predictions by averaging the outcomes of each tree. The RF is preferred due to its ability to learn intricate non-linear dependencies and tolerance to data heterogeneity. The significance of features with respect to the output can be determined using GI. To obtain the optimal feature set while managing computational load, the backward feature elimination method was utilized. The process began with a model using the complete set of *n* features. Subsequently, all *n*-1 combinations of features were systematically tested, and the next set was constructed in each iteration by eliminating the least significant feature. The model performance at each cycle was determined based on the RMSE of the test set. The process terminates when the RMSE reaches its minimum, just before exhibiting an increase rate of 10 % in the next iteration.

**Table 1**. List of 30 features used in this work. The subscripts (superscripts) of M, A, C, Tot, and H denote the mobile ions (Li, Na), anion, cation, total atoms, and hybridization.

| Features | Description |
|---|---|
| $<\omega>_M, <\omega>_A, <\omega>_{Tot}$ | Phonon band center of mobile ion, anion, and total atoms |
| $f_{max}^M, f_{max}^A, f_{max}^H$ | Frequency at maximum phonon DOS of mobile ion, anion, and hybridization of mobile ion and anion below 30 meV |
| $P_{max}^M, P_{max}^A, P_{max}^H$ | Maximum phonon DOS below 30 meV per corresponding atoms |
| $f_{img}$ | The existence of soft mode |
| $R_{M/Tot}, R_{A/Tot}, R_{H/Tot}$ | The ratio of atom-projected phonon DOS per total phonon DOS below 30 meV |
| $S$ | Vibrational entropy at RT |
| $E_G$ | Energy band gap |
| $\Delta H_f$ | Heat of formation |
| $V_{f.u}, V_M, V_A, V_{Tot}$ | Volume per formula units, mobile ion, anion, and total atoms |
| $d_{M-A}, d_{M-M}$ | The minimum distance between Li (Na) and anion, Li (Na) and Li (Na) |
| $N_{f.u}$ | Number of formula units in the conventional cell |
| $\rho$ | Density |
| $m$ | Total mass |
| $N_M$ | Number of mobile ions in formula unit |
| M-disorder, A-disorder, C-disorder, Tot-disorder | Site atomic disorder of mobile ion, anion, cation, and total atoms |

Firstly, we compiled a list of 45 Li and Na-ion conductors along with their structures and ionic conductivity at RT, denoted as σ, as detailed in Table S1. Figure S1 shows the histogram of the dataset categorized by log(σ). Feature selection is one of the most important parts to improve a model performance. To consider multifaceted perspectives, we selected features incorporating phonon, structural, and electronic properties based on previous literatures.[27,29,66–70] and physical intuition. We also defined several features to quantitatively describe specific properties related to phonon DOS and disorders. Table 1 summarizes the selected 30 features.

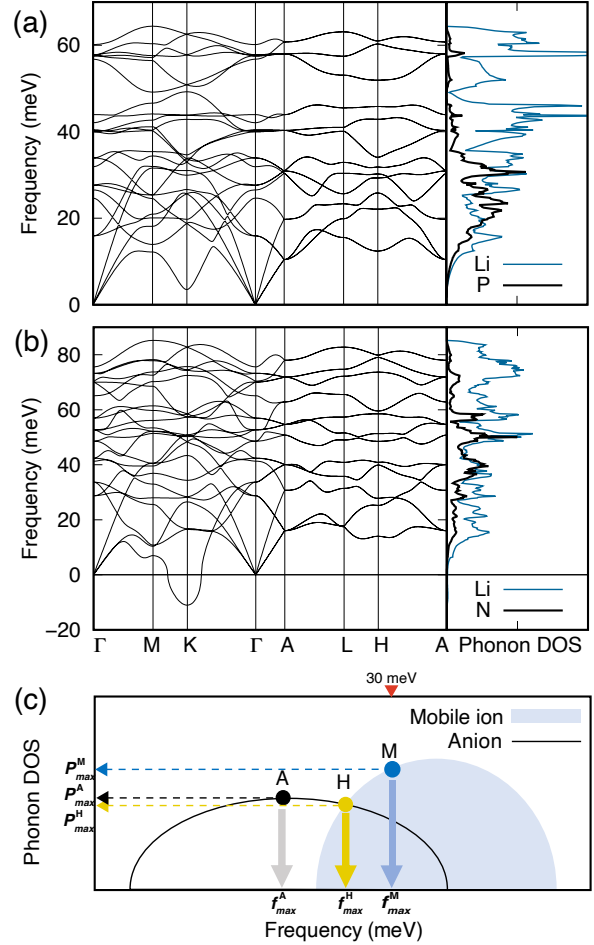

**Figure 2.** Phonon dispersions and density of states (DOS) of (a) Li$_3$P and (b) β-Li$_3$N as representatives. The negative phonon frequency indicates the phonon soft mode. (c) Schematic phonon DOS. $P_{max}^M$, $P_{max}^A$, and $P_{max}^H$ indicate the maximum phonon DOS below 30 meV for mobile ion (Li or Na), anion, and their intersection (hybridization) region. $f_{max}^M$, $f_{max}^A$, and $f_{max}^H$ are the corresponding frequencies to $P_{max}^M$, $P_{max}^A$, and $P_{max}^H$, respectively.

To consider features related to lattice dynamics, we analyze phonon bands and include various representations of phonon characteristics. Figure 2(a) and (b) show the phonon dispersion curves of Li$_3$P ($\sigma$ = 1.0 × 10$^{-3}$ S/cm) and β-Li$_3$N ($\sigma$ = 2.085 × 10$^{-4}$ S/cm), respectively, as representative examples. First, we used <ω> to denote the average vibrational frequency so-called 'phonon band center' following the previous study.[29] The quantity of <ω> is expressed as $<\omega> = \frac{\int \omega \times \text{DOS}(\omega) d\omega}{\int \text{DOS}(\omega) d\omega}$. <ω>$_M$, <ω>$_A$, and <ω>$_{Tot}$ correspond to the phonon band centers of mobile ion (Li or Na), anion, and total atoms, respectively.

In previous studies, the glassy Li$_3$P$_4$ phase has been observed to exhibit a lower peak of Li and larger overlap with anion spectra than crystalline γ-Li$_3$P$_4$.[66] These characteristics were thought to enhance ion mobility at low temperatures by promoting the paddlewheel dynamics. However, other mechanisms to enhance ion mobility related to such phonon spectra feature have been proposed recently,[31,71] such as anharmonic phonon coupling. We constructed features that relate to the phonon peak position and defined a feature associated with the overlapping phonon DOS of Li (Na) and anion. Specifically, the features, $f_{max}^M$ and $f_{max}^A$ correspond to the frequency at the maximum phonon DOS of mobile ion and anion, respectively. The frequency at the phonon peak for 'hybridization', *i.e.*, the overlap region of mobile ion and anion, is denoted as $f_{max}^H$ as shown in Figure 2(c). $P_{max}^M$, $P_{max}^A$, and $P_{max}^H$ are defined as phonon DOS values at $f_{max}^M$, $f_{max}^A$, and $f_{max}^H$ per corresponding atoms.

In addition, we restricted the highest phonon frequency to 30 meV, which is close to the RT energy scale of ~25 meV. Phonons with energies higher than the RT energy of 25 meV have significantly smaller average number of phonons at RT because the average number of phonons, $\langle n \rangle$ at fixed temperature follows the Planck distribution, $\langle n \rangle = \frac{1}{\exp(\frac{\hbar\omega}{k_BT}-1)}$. This restriction also emphasizes the low frequency region that is related to lattice softness.

We further selected features that can be associated with lattice softness. The feature $f_{img}$ denotes the existence of soft mode in the phonon bands since it can be considered a precursor to ionic diffusion.[27] We designated a value of 0 if there are no imaginary frequencies. Values 1 and 2 were assigned if soft modes originate from mobile and other ions, respectively. Furthermore, we considered additional phonon-related properties such as the phonon DOS of mobile ion, anion, and hybridization region per total phonon DOS ($R_{M/Tot}$, $R_{A/Tot}$, $R_{H/Tot}$) below 30 meV, along with vibrational entropy, S at RT.[70] The ratio of the phonon DOS below 30 meV indicates how the phonon DOS is relatively distributed in the low frequency region. The softer lattice tends to exhibit higher vibrational entropy.[70]

For electronic and structural properties, we selected features, similar to previous literature,[47,48] such as band gap ($E_G$), the heat of formation ($\Delta H_f$), volume per formula units, mobile ion, anion, and total atoms ($V_{f.u}$, $V_M$, $V_A$, $V_{Tot}$), the minimum distance between Li (Na) and anion ($d_{M-A}$), Li (Na) and Li (Na) ($d_{M-M}$), and the number of formula units ($N_{f.u}$) in the conventional cell. In addition, we included the properties derived from the chemical composition of materials such as density ($\rho$),[46] mass ($m$), and the number of mobile ions in the formula unit cell ($N_M$).

We also incorporated site-disorder properties into the feature set, as activation energies for Li-ion diffusion are expected to decrease with increasing site-disorder.[67–69] We quantified disorders using site occupancy and the Wyckoff site number. M-disorder, A-disorder, C-disorder, and Tot-disorder denote the amount of the mobile ion, anion, cation, and total site-disorder in the given structure, respectively. Computational details for quantifying the disorder in an example system are described in the Supporting information.

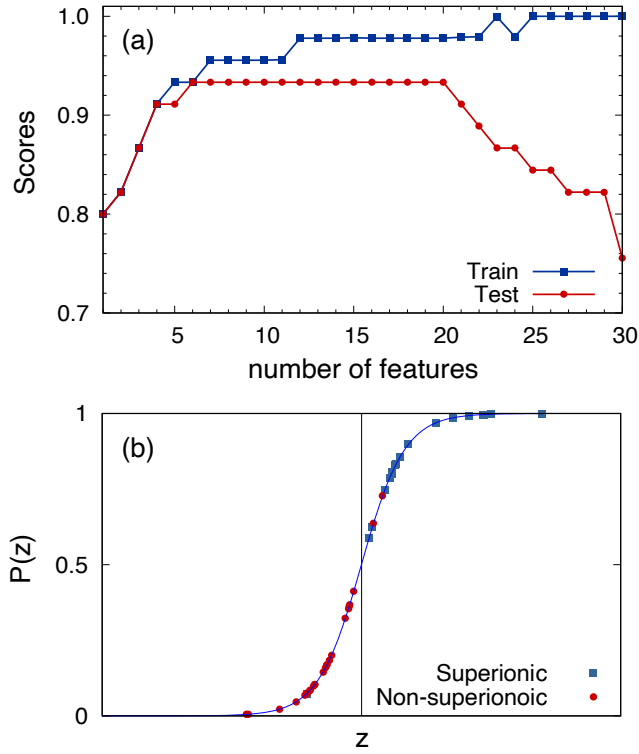

**Figure 3**. Prediction performance of the LR models. (a) The highest accuracy scores according to the number of features. (b) The classification performance using 6 features with an accuracy of 0.933 and an F1 score of 0.927. The blue squares and red dots represent the superionic and non-superionic materials based on experimental observations, respectively.

In the LR approach for the binary classification model, materials were categorized as superionic and non-superionic based on a threshold value of $10^{-4}$ S/cm, and classified into binary outputs of 1 ($\log(\sigma) > -4$ Scm$^{-1}$) and 0 ($\log(\sigma) \leq -4$ Scm$^{-1}$) as suggested by the previous study.[48] To search for the best combination of predictors, we evaluated the prediction performance by considering all possible combinations of the 30 primary features. Figure 3(a) shows the highest scores of our models depending on the number of features. The score of the test set increases sharply up to a score of 0.933 and remains constant until 20 features. The performance of the test set deteriorates with more than 20 features, unlike the training set displaying high accuracy, indicating overfitting. Using a single feature set, the best model achieves an accuracy of 0.8 with the $N_{f.u}$ feature. Combinations with other features further enhance performance, reaching up to 0.933 with 6 features.

Table 2 shows the best predictors with 2 to 6 combinatorial features and their accuracies. Even with just four features, the prediction performance presents an accuracy greater than 0.9, as depicted in Figure 3(a). The best combinations of 4 features, scoring an accuracy of 0.911, include $N_{f.u}$, $R_{M/Tot}$, C-disorder, and $N_M$ ($V_{f.u}$). $N_{f.u}$, $R_{M/Tot}$, and C-disorder consistently appear in the best combinations ranging from 4 to 19 features. The combination of $R_{H/Tot}$, S, and A-disorder, together with $N_{f.u}$, $R_{M/Tot}$, and C-disorder exhibit the highest performance with an accuracy of 0.933. Conversely, combinations of 4 and 6 features excluding phonon-related attributes result in accuracies of 0.867 and 0.889, respectively. These results suggest the influential role of phonon characteristics in enhancing the performance of the ML model.

**Table 2.** The optimal feature set and its corresponding test accuracy with respect to the number of features.

| Number of features | Best Combinations | Accuracy |
|---|---|---|
| 2 | $N_{f.u}$, C-disorder<br>$N_{f.u}$, A-disorder<br>$N_{f.u}$, Tot-disorder | 0.822 |
| 3 | $N_{f.u}$, C-disorder, $E_G$<br>$N_{f.u}$, C-disorder, $N_M$<br>$N_{f.u}$, $R_{M/Tot}$, $f_{img}$,<br>$N_{f.u}$, $f_{max}^A$, S | 0.867 |
| 4 | $N_{f.u}$, $R_{M/Tot}$, C-disorder, $N_M$<br>$N_{f.u}$, $R_{M/Tot}$, C-disorder, $V_{f.u}$ | 0.911 |
| 5 | $N_{f.u}$, $R_{M/Tot}$, C-disorder, A-disorder, S<br>$N_{f.u}$, $R_{M/Tot}$, C-disorder, $d_{M-M}$, $m$ | 0.911 |
| 6 | $N_{f.u}$, $R_{M/Tot}$, $R_{H/Tot}$, C-disorder, A-disorder, S | 0.933 |

Figure 3(b) illustrates the logistic function $P(z)$ obtained using the 6 features of $N_{f.u}$, $R_{M/Tot}$, $R_{H/Tot}$, C-disorder, A-disorder, and S. A value of $P(z)$ greater than (less than) 0.5 indicates the prediction of superionic (non-superionic) behavior. Out of the 45 materials, three are misclassified: $Li_4BN_3H_{10}$, $Li_6PS_5I$, and $Li_2CaN_2H_2$. The confusion matrix is shown in Figure S2.

$$z = -0.427\, N_{f.u} + 4.523\, R_{M/Tot} + 2.561\, R_{H/tot} + 1.305\, C\text{-disorder} +$$
$$2.220\, A\text{-disorder} + 0.007\, S - 2.552 \quad (1)$$

Equation (1) presents the $z$ function for the best combinations of 6 features in Table 2. By examining the signs of the coefficients $w_i$ in the $z$ function, we can infer their influence on the likelihood of superionic behavior. From a structural perspective, crystal structures with fewer number of formula units tend to exhibit high ionic conductivity ($w_{N_{f.u}} < 0$). The presence of cation and anion disorders positively influences fast ionic transport ($w_{\text{C-disorder}} > 0$, $w_{\text{A-disorder}} > 0$). Regarding dynamic properties, softened lattice vibrations of mobile ions (Li or Na) contribute to the increased probability of superionic conductivity ($w_{R_{M/Tot}} > 0$). A larger overlap of mobile ion and anion phonon DOS promotes superionic conductor ($w_{R_{H/Tot}} > 0$). Furthermore, higher vibrational entropy that is associated with the softness of the lattice enhances the fast ionic conduction ($w_S > 0$). The Pearson correlation coefficients of $N_{f.u}$, $R_{M/Tot}$, $R_{H/Tot}$, C-disorder, A-disorder, and S to the ionic conductivity are -0.400, 0.357, 0.296, 0.349, 0.238, and 0.493, respectively, indicating consistent positive/negative relationships with the LR results (See Table S3).

Table 3 presents the GI derived from the RF algorithm, which evaluates the significance of each feature on the output. Among these six features, $R_{H/Tot}$, $R_{M/Tot}$, and $N_{f.u}$ are the most important features for classifying superionic conductors, each with similar GIs of > 0.2. Following closely is vibrational entropy with a GI of 0.146. Conversely, disorder-related features contribute minimally to the classification of ionic conductivity. It is noteworthy that features related to phonons exhibit high GI values, showing their relevance to fast ion transport.

Table 3. The Gini importance of each feature from the best LR and RF model.

| Rank | Classification | | Regression | |
|---|---|---|---|---|
| 1 | $R_{H/Tot}$ | 0.236 | $f_{max}^{A}$ | 0.312 |
| 2 | $R_{M/Tot}$ | 0.235 | $P_{max}^{M}$ | 0.233 |
| 3 | $N_{f.u}$ | 0.227 | $N_{f.u}$ | 0.216 |
| 4 | S | 0.146 | $N_M$ | 0.174 |
| 5 | C-disorder | 0.092 | C-disorder | 0.065 |
| 6 | A-disorder | 0.064 | | |

We further developed the ensemble regression model of RF to predict the ionic conductivity and assess the contribution of each feature to estimating it. Figure 4(a) shows the prediction performance with RMSE and $R^2$ values according to the number of features. Each point represents the RMSE and $R^2$ obtained from the best combination of features using the backward feature elimination method. Among models with a single feature, the best performance is achieved when using $f_{max}^{A}$, with an RMSE of 1.696 S/cm and $R^2$ of 0.296, respectively. As the number of features increases, the RMSE values decrease to 1.179 S/cm and then increase with more than 5 features, reaching up to 1.645 S/cm with 30 features. The combination of five features, $f_{max}^{A}$, $P_{max}^{M}$, $N_{f.u}$, $N_M$, and C-disorder, demonstrates the best prediction performance an RMSE of 1.179 S/cm and $R^2$ of 0.710, as shown in Figure 4(b). The best performance of the RF models is not better than the classification model using LR despite a high $R^2$ of 0.96 in the training set. This would be because accurately predicting ionic conductivity is more complex than classifying properties.

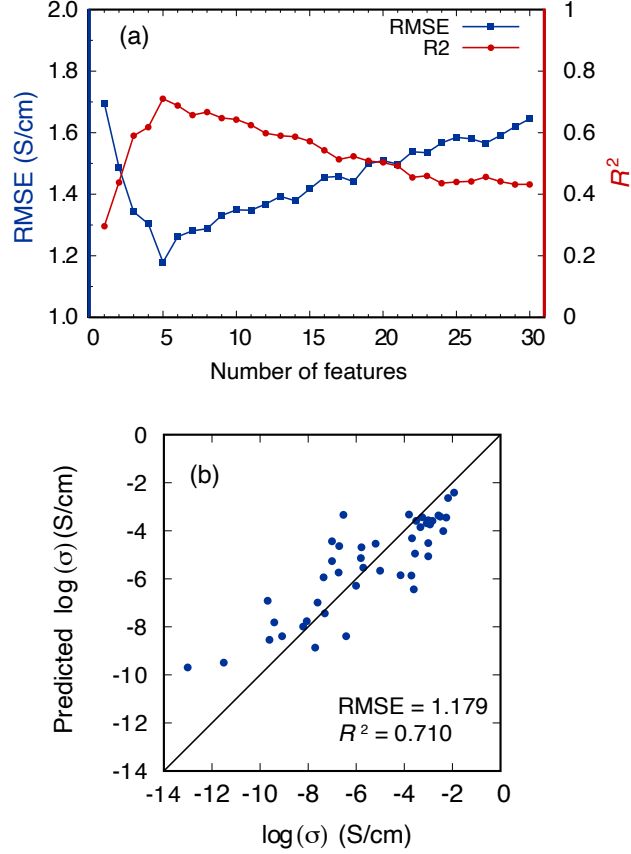

**Figure 4.** Prediction performance of the RF. (a) RMSE and $R^2$ depending on the number of features. (b) Comparison of the predicted and experimental ionic conductivity log ($\sigma$).

Table 3 represents the GIs of $f_{max}^A$, $P_{max}^M$, $N_{f.u}$, $N_M$, and C-disorder. The frequency at maximum phonon DOS of anion below 30 meV, $f_{max}^A$, is the highest-ranked feature with a GI of 0.312. The $P_{max}^M$ and $N_{f.u}$ rank second and third, contributing comparably to this model, with similar GIs of 0.233 and 0.216, respectively. Mobile ion density ($N_M$) is the fourth important feature, while C-disorder is the least important in predicting ionic conductivity.

Note that two phonon-related features ($f_{max}^A$, $P_{max}^M$) are included as important features in the best performance regression model. Like the classification model, the most significant properties in the regression model are associated with the phonon properties according to the GIs. Considering the dynamic attributes of both mobile ions and anions improves the performance of identifying materials with high ionic conductivity. These results are consistent with the previous literature reporting the correlation with lattice dynamics on ionic transport,[6,24–31] and our results further show their statistical significance through ML analysis. $N_{f.u}$ and C-disorder features are also included in the best feature combination as in the LR model, indicating their role in determining ionic conductivity. As shown in the GI and $z$ function, the positive influence of disorder-related features is consistent with previous studies.[67–69]

**Table 4.** Candidates for superionic conductors screened by the LR model.

| MP ID | Composition | Space group |
|---|---|---|
| 4527 | $Li_8SnO_6$ | R-3 |
| 6844 | $KLi_3Si_{12}Sn_2O_{30}$ | P6/mcc |
| 8610 | $Li_8PtO_6$ | R-3 |
| 14364 | $Cs_2LiAsO_4$ | $Cmc2_1$ |
| 16055 | $KLi_3Zr_2Si_{12}O_{30}$ | P6/mcc |
| 17208 | $K_2Li_{14}Zr_3O_{14}$ | Immm |
| 18711 | $LiNdP_4O_{12}$ | C2/c |
| 504806 | $K_2Li_{14}Pb_3O_{14}$ | Immm |
| 504810 | $Rb_2Li_{14}Pb_3O_{14}$ | Immm |
| 774749 | $RbNa_3Li_{12}Ti_4O_{16}$ | I4/m |
| 753546 | $Li_8TiS_6$ | P63cm |

Finally, we apply our LR model to search for new SSE candidates. All structures and phonon data are acquired from the MP[55] and phonon database at Kyoto University.[56–58] We have screened all the Li-based materials from the phonon database with the S or O anions. The materials that have a small gap (< 1.0 eV) are eliminated to minimize the electronic conductions as electrolytes. We used the materials that satisfy the convex hull energy ($E_{hull}$) is 0 eV/atom to prevent sample decomposition. From these requirements and restrictions, we compiled 259 Li-contained materials and added additional 5 materials (LiErSe$_2$, LiHo$_3$Ge$_2$O$_8$F$_2$, LiBiF$_4$, LiCaGaF$_6$, LiInF$_4$) from the previous MD report[49] for comparison.

By screening 264 datasets based on our LR model with six features ($N_{f.u}$, $R_{M/Tot}$, $R_{H/Tot}$, C-disorder, A-disorder, S), we identified 11 materials out of 264 (~4%) that may exhibit superionic characters as listed in Table 4. Among the predicted candidates, Li$_8$SnO$_6$ has been explored as an electrode material for LIBs due to its high lithium content,[72–74] and DFT studies reported the migration energy ranging from 0.20 eV to 1.06 eV across different paths and methods.[72,73] Other candidates listed in Table 4 have not yet been identified through either DFT-MD or experiments to the best of our knowledge. To further validate the performance of our model, we compared the results to the previous report using the DFT-MD simulations.[49] Our prediction matches the calculated ionic conductivity with 87.0 % accuracy as shown in Table S4. Although we observed a few false negative results compared to the previous MD results, we anticipate that our screening model with phonon features has prediction power and will facilitate the discovery of superionic conductors.

In conclusion, we have developed ML models utilizing LR and RF algorithms to predict the ionic conductivity of SSE materials. Our approach involved careful selection and quantification of features from both dynamic and static properties using the DFT calculations and experimental structures. The LR model exhibits an accuracy of 93 % with six features: $N_{f.u}$, $R_{M/Tot}$, $R_{H/Tot}$, C-disorder, A-disorder, and S. A structure with a small number of formula units, lattice softness around Li (Na) ions, a large overlap of phonon DOS in mobile ions and anions, high disorder of cations and anions, and high vibrational entropy promote superionic behavior. The RF model presents the prediction performance with an RMSE of 1.179 S/m and an $R^2$ of 0.710. Notably, phonon-related features such as $f_{max}^A$ and $P_{max}^M$, along with structure-related features including $N_{f.u}$, $N_M$, and C-disorder, play crucial roles in predicting ionic conductivity in the RF regression model. In both classification and regression models, phonon-related features exhibit higher GI, quantitatively indicating their strong relevance to ionic conductivity. Thus, our ML models demonstrate the significance of dynamic properties in predicting and understanding the ionic conductivity of materials. Finally, based on the developed models, we screened the 264 Li-based materials for SSE materials and suggested 11 candidates for superionic conductors. We believe that our understanding can guide the future design of advanced superionic conductors.

**SUPPORTING INFORMATION**

List of the training set, which includes composition, ionic conductivity at RT, and space group of materials, along with histograms of the dataset, an example for the disorder features, the confusion matrix, the Pearson correlation map of the features and conductivity, and ML screened results for comparison to previous MD study.

**ACKNOWLEDGEMENTS**

We thank to Jaewook Kim and Heungsik Kim for their helpful discussions. This work was supported by the National Research Foundation of Korea (NRF) (Grant No. 2022R1F1A1063011, 2021R1A4A1029780) and KISTI Supercomputing Center (Project No. KSC-2023-CRE-0395). This work was also supported by Learning & Academic research institution for Master's·PhD students, and Postdocs (LAMP) Program of the NRF grant funded by the Ministry of Education (No. RS-2023-00301914).

# REFERENCES

(1) Armand, M.; Tarascon, J.-M. Building Better Batteries. *Nature* **2008**, *451* (7179), 652–657. https://doi.org/10.1038/451652a.

(2) Dunn, B.; Kamath, H.; Tarascon, J.-M. Electrical Energy Storage for the Grid: A Battery of Choices. *Science (80-. ).* **2011**, *334* (6058), 928–935. https://doi.org/10.1126/science.1212741.

(3) Tarascon, J.-M.; Armand, M. Issues and Challenges Facing Rechargeable Lithium Batteries. *Nature* **2001**, *414* (6861), 359–367. https://doi.org/10.1038/35104644.

(4) Chen, Y.; Kang, Y.; Zhao, Y.; Wang, L.; Liu, J.; Li, Y.; Liang, Z.; He, X.; Li, X.; Tavajohi, N.; Li, B. A Review of Lithium-Ion Battery Safety Concerns: The Issues, Strategies, and Testing Standards. *J. Energy Chem.* **2021**, *59*, 83–99. https://doi.org/10.1016/j.jechem.2020.10.017.

(5) Wang, Q.; Jiang, L.; Yu, Y.; Sun, J. Progress of Enhancing the Safety of Lithium Ion Battery from the Electrolyte Aspect. *Nano Energy* **2019**, *55*, 93–114. https://doi.org/10.1016/j.nanoen.2018.10.035.

(6) Bachman, J. C.; Muy, S.; Grimaud, A.; Chang, H. H.; Pour, N.; Lux, S. F.; Paschos, O.; Maglia, F.; Lupart, S.; Lamp, P.; Giordano, L.; Shao-Horn, Y. Inorganic Solid-State Electrolytes for Lithium Batteries: Mechanisms and Properties Governing Ion Conduction. *Chem. Rev.* **2016**, *116* (1), 140–162. https://doi.org/10.1021/acs.chemrev.5b00563.

(7) Zhao, W.; Yi, J.; He, P.; Zhou, H. Solid-State Electrolytes for Lithium-Ion Batteries: Fundamentals, Challenges and Perspectives. *Electrochem. Energy Rev.* **2019**, *2* (4), 574–605. https://doi.org/10.1007/s41918-019-00048-0.

(8) Zheng, Y.; Yao, Y.; Ou, J.; Li, M.; Luo, D.; Dou, H.; Li, Z.; Amine, K.; Yu, A.; Chen, Z.

A Review of Composite Solid-State Electrolytes for Lithium Batteries: Fundamentals, Key Materials and Advanced Structures. *Chem. Soc. Rev.* **2020**, *49* (23), 8790–8839. https://doi.org/10.1039/D0CS00305K.

(9) Li, J.; Ma, C.; Chi, M.; Liang, C.; Dudney, N. J. Solid Electrolyte: The Key for High-Voltage Lithium Batteries. *Adv. Energy Mater.* **2015**, *5* (4), 1401408. https://doi.org/10.1002/aenm.201401408.

(10) Knauth, P. Inorganic Solid Li Ion Conductors: An Overview. *Solid State Ionics* **2009**, *180* (14–16), 911–916. https://doi.org/10.1016/j.ssi.2009.03.022.

(11) Li, W.; Chen, L.; Sun, Y.; Wang, C.; Wang, Y.; Xia, Y. All-Solid-State Secondary Lithium Battery Using Solid Polymer Electrolyte and Anthraquinone Cathode. *Solid State Ionics* **2017**, *300*, 114–119. https://doi.org/10.1016/j.ssi.2016.12.013.

(12) Janek, J.; Zeier, W. G. A Solid Future for Battery Development. *Nat. Energy* **2016**, *1* (9), 16141. https://doi.org/10.1038/nenergy.2016.141.

(13) Kamaya, N.; Homma, K.; Yamakawa, Y.; Hirayama, M.; Kanno, R.; Yonemura, M.; Kamiyama, T.; Kato, Y.; Hama, S.; Kawamoto, K.; Mitsui, A. A Lithium Superionic Conductor. *Nat. Mater.* **2011**, *10* (9), 682–686. https://doi.org/10.1038/nmat3066.

(14) Aono, H.; Sugimoto, E.; Sadaoka, Y.; Imanaka, N.; Adachi, G. Ionic Conductivity of Solid Electrolytes Based on Lithium Titanium Phosphate. *J. Electrochem. Soc.* **1990**, *137* (4), 1023–1027. https://doi.org/10.1149/1.2086597.

(15) Murugan, R.; Thangadurai, V.; Weppner, W. Fast Lithium Ion Conduction in Garnet-Type $Li_7La_3Zr_2O_{12}$. *Angew. Chemie Int. Ed.* **2007**, *46* (41), 7778–7781. https://doi.org/10.1002/anie.200701144.

(16) Inaguma, Y.; Liquan, C.; Itoh, M.; Nakamura, T.; Uchida, T.; Ikuta, H.; Wakihara, M.

High Ionic Conductivity in Lithium Lanthanum Titanate. *Solid State Commun.* **1993**, *86* (10), 689–693. https://doi.org/10.1016/0038-1098(93)90841-A.

(17) Rao, R. P.; Adams, S. Studies of Lithium Argyrodite Solid Electrolytes for All-Solid-State Batteries. *Phys. Status Solidi Appl. Mater. Sci.* **2011**, *208* (8), 1804–1807. https://doi.org/10.1002/pssa.201001117.

(18) Pervez, S. A.; Cambaz, M. A.; Thangadurai, V.; Fichtner, M. Interface in Solid-State Lithium Battery: Challenges, Progress, and Outlook. *ACS Appl. Mater. Interfaces* **2019**, *11* (25), 22029–22050. https://doi.org/10.1021/acsami.9b02675.

(19) Sheng, O.; Jin, C.; Ding, X.; Liu, T.; Wan, Y.; Liu, Y.; Nai, J.; Wang, Y.; Liu, C.; Tao, X. A Decade of Progress on Solid-State Electrolytes for Secondary Batteries: Advances and Contributions. *Adv. Funct. Mater.* **2021**, *31* (27), 2100891. https://doi.org/10.1002/adfm.202100891.

(20) Martínez-Juárez, A.; Pecharromán, C.; Iglesias, J. E.; Rojo, J. M. Relationship between Activation Energy and Bottleneck Size for $Li^+$ Ion Conduction in NASICON Materials of Composition $LiMM'(PO_4)_3$; M, M′ = Ge, Ti, Sn, Hf. *J. Phys. Chem. B* **1998**, *102* (2), 372–375. https://doi.org/10.1021/jp973296c.

(21) Krauskopf, T.; Culver, S. P.; Zeier, W. G. Bottleneck of Diffusion and Inductive Effects in $Li_{10}Ge_{1-X}Sn_xP_2S_{12}$. *Chem. Mater.* **2018**, *30* (5), 1791–1798. https://doi.org/10.1021/acs.chemmater.8b00266.

(22) Avdeev, M.; Sale, M.; Adams, S.; Rao, R. P. Screening of the Alkali-Metal Ion Containing Materials from the Inorganic Crystal Structure Database (ICSD) for High Ionic Conductivity Pathways Using the Bond Valence Method. *Solid State Ionics* **2012**, *225*, 43–46. https://doi.org/10.1016/j.ssi.2012.02.014.

(23) Wang, Y.; Richards, W. D.; Ong, S. P.; Miara, L. J.; Kim, J. C.; Mo, Y.; Ceder, G. Design Principles for Solid-State Lithium Superionic Conductors. *Nat. Mater.* **2015**, *14* (10), 1026–1031. https://doi.org/10.1038/nmat4369.

(24) Muy, S.; Schlem, R.; Shao-Horn, Y.; Zeier, W. G. Phonon–Ion Interactions: Designing Ion Mobility Based on Lattice Dynamics. *Adv. Energy Mater.* **2021**, *11* (15), 2002787. https://doi.org/10.1002/aenm.202002787.

(25) Wakamura, K. Roles of Phonon Amplitude and Low-Energy Optical Phonons on Superionic Conduction. *Phys. Rev. B* **1997**, *56* (18), 11593–11599. https://doi.org/10.1103/PhysRevB.56.11593.

(26) Li, X.; Benedek, N. A. Enhancement of Ionic Transport in Complex Oxides through Soft Lattice Modes and Epitaxial Strain. *Chem. Mater.* **2015**, *27* (7), 2647–2652. https://doi.org/10.1021/acs.chemmater.5b00445.

(27) Gupta, M. K.; Goel, P.; Mittal, R.; Choudhury, N.; Chaplot, S. L. Phonon Instability and Mechanism of Superionic Conduction in $Li_2O$. *Phys. Rev. B* **2012**, *85* (18), 184304. https://doi.org/10.1103/PhysRevB.85.184304.

(28) Krauskopf, T.; Pompe, C.; Kraft, M. A.; Zeier, W. G. Influence of Lattice Dynamics on $Na^+$ Transport in the Solid Electrolyte $Na_3PS_{4-x}Se_x$. *Chem. Mater.* **2017**, *29* (20), 8859–8869. https://doi.org/10.1021/acs.chemmater.7b03474.

(29) Muy, S.; Bachman, J. C.; Giordano, L.; Chang, H. H.; Abernathy, D. L.; Bansal, D.; Delaire, O.; Hori, S.; Kanno, R.; Maglia, F.; Lupart, S.; Lamp, P.; Shao-Horn, Y. Tuning Mobility and Stability of Lithium Ion Conductors Based on Lattice Dynamics. *Energy Environ. Sci.* **2018**, *11* (4), 850–859. https://doi.org/10.1039/c7ee03364h.

(30) Sagotra, A. K.; Chu, D.; Cazorla, C. Influence of Lattice Dynamics on Lithium-Ion

Conductivity: A First-Principles Study. *Phys. Rev. Mater.* **2019**, *3* (3), 035405. https://doi.org/10.1103/PhysRevMaterials.3.035405.

(31) Xu, Z.; Chen, X.; Zhu, H.; Li, X. Anharmonic Cation–Anion Coupling Dynamics Assisted Lithium-Ion Diffusion in Sulfide Solid Electrolytes. *Adv. Mater.* **2022**, *34* (49), 2207411. https://doi.org/10.1002/adma.202207411.

(32) Muy, S.; Voss, J.; Schlem, R.; Koerver, R.; Sedlmaier, S. J.; Maglia, F.; Lamp, P.; Zeier, W. G.; Shao-Horn, Y. High-Throughput Screening of Solid-State Li-Ion Conductors Using Lattice-Dynamics Descriptors. *iScience* **2019**, *16*, 270–282. https://doi.org/10.1016/j.isci.2019.05.036.

(33) Deng, Z.; Radhakrishnan, B.; Ong, S. P. Rational Composition Optimization of the Lithium-Rich $Li_3OCl_{1-x}Br_x$ Anti-Perovskite Superionic Conductors. *Chem. Mater.* **2015**, *27* (10), 3749–3755. https://doi.org/10.1021/acs.chemmater.5b00988.

(34) Kang, J.; Han, B. First-Principles Characterization of the Unknown Crystal Structure and Ionic Conductivity of $Li_7P_2S_8I$ as a Solid Electrolyte for High-Voltage Li Ion Batteries. *J. Phys. Chem. Lett.* **2016**, *7* (14), 2671–2675. https://doi.org/10.1021/acs.jpclett.6b01050.

(35) Lacivita, V.; Artrith, N.; Ceder, G. Structural and Compositional Factors That Control the Li-Ion Conductivity in LiPON Electrolytes. *Chem. Mater.* **2018**, *30* (20), 7077–7090. https://doi.org/10.1021/acs.chemmater.8b02812.

(36) You, D.; Zhang, H.; Ganorkar, S.; Kim, T.; Schroers, J.; Vlassak, J. J.; Lee, D. Electrical Resistivity as a Descriptor for Classification of Amorphous versus Crystalline Phases of Alloys. *Acta Mater.* **2022**, *231*, 117861. https://doi.org/10.1016/j.actamat.2022.117861.

(37) You, D.; Ganorkar, S.; Kim, S.; Kang, K.; Shin, W.-Y.; Lee, D. Machine Learning-Based Prediction Models for Formation Energies of Interstitial Atoms in HCP Crystals. *Scr.*

*Mater.* **2020**, *183*, 1–5. https://doi.org/10.1016/j.scriptamat.2020.02.042.

(38) Song, J.; Jo, H.; Kim, T.; Lee, D. Experimental Data Management Platform for Data-Driven Investigation of Combinatorial Alloy Thin Films. *APL Mater.* **2023**, *11* (9), 09117. https://doi.org/10.1063/5.0162158.

(39) Lee, D.; You, D.; Lee, D.; Li, X.; Kim, S. Machine-Learning-Guided Prediction Models of Critical Temperature of Cuprates. *J. Phys. Chem. Lett.* **2021**, *12* (26), 6211–6217. https://doi.org/10.1021/acs.jpclett.1c01442.

(40) Lee, Y.-L.; Lee, H.; Kim, T.; Byun, S.; Lee, Y. K.; Jang, S.; Chung, I.; Chang, H.; Im, J. Data-Driven Enhancement of ZT in SnSe-Based Thermoelectric Systems. *J. Am. Chem. Soc.* **2022**, *144* (30), 13748–13763. https://doi.org/10.1021/jacs.2c04741.

(41) Allam, O.; Cho, B. W.; Kim, K. C.; Jang, S. S. Application of DFT-Based Machine Learning for Developing Molecular Electrode Materials in Li-Ion Batteries. *RSC Adv.* **2018**, *8* (69), 39414–39420. https://doi.org/10.1039/C8RA07112H.

(42) Choi, E.; Jo, J.; Kim, W.; Min, K. Searching for Mechanically Superior Solid-State Electrolytes in Li-Ion Batteries via Data-Driven Approaches. *ACS Appl. Mater. Interfaces* **2021**, *13* (36), 42590–42597. https://doi.org/10.1021/acsami.1c07999.

(43) Hu, S.; Huang, C. Machine-Learning Approaches for the Discovery of Electrolyte Materials for Solid-State Lithium Batteries. *Batteries* **2023**, *9* (4), 228. https://doi.org/10.3390/batteries9040228.

(44) Zhang, Y.; He, X.; Chen, Z.; Bai, Q.; Nolan, A. M.; Roberts, C. A.; Banerjee, D.; Matsunaga, T.; Mo, Y.; Ling, C. Unsupervised Discovery of Solid-State Lithium Ion Conductors. *Nat. Commun.* **2019**, *10* (1), 5260. https://doi.org/10.1038/s41467-019-13214-1.

(45) Hargreaves, C. J.; Gaultois, M. W.; Daniels, L. M.; Watts, E. J.; Kurlin, V. A.; Moran, M.; Dang, Y.; Morris, R.; Morscher, A.; Thompson, K.; Wright, M. A.; Prasad, B.-E.; Blanc, F.; Collins, C. M.; Crawford, C. A.; Duff, B. B.; Evans, J.; Gamon, J.; Han, G.; Leube, B. T.; Niu, H.; Perez, A. J.; Robinson, A.; Rogan, O.; Sharp, P. M.; Shoko, E.; Sonni, M.; Thomas, W. J.; Vasylenko, A.; Wang, L.; Rosseinsky, M. J.; Dyer, M. S. A Database of Experimentally Measured Lithium Solid Electrolyte Conductivities Evaluated with Machine Learning. *npj Comput. Mater.* **2023**, *9* (1), 9. https://doi.org/10.1038/s41524-022-00951-z.

(46) Cubuk, E. D.; Sendek, A. D.; Reed, E. J. Screening Billions of Candidates for Solid Lithium-Ion Conductors: A Transfer Learning Approach for Small Data. *J. Chem. Phys.* **2019**, *150* (21), 214701. https://doi.org/10.1063/1.5093220.

(47) Jalem, R.; Nakayama, M.; Kasuga, T. An Efficient Rule-Based Screening Approach for Discovering Fast Lithium Ion Conductors Using Density Functional Theory and Artificial Neural Networks. *J. Mater. Chem. A* **2014**, *2* (3), 720–734. https://doi.org/10.1039/c3ta13235h.

(48) Sendek, A. D.; Yang, Q.; Cubuk, E. D.; Duerloo, K. A. N.; Cui, Y.; Reed, E. J. Holistic Computational Structure Screening of More than 12 000 Candidates for Solid Lithium-Ion Conductor Materials. *Energy Environ. Sci.* **2017**, *10* (1), 306–320. https://doi.org/10.1039/c6ee02697d.

(49) Sendek, A. D.; Cubuk, E. D.; Antoniuk, E. R.; Cheon, G.; Cui, Y.; Reed, E. J. Machine Learning-Assisted Discovery of Solid Li-Ion Conducting Materials. *Chem. Mater.* **2019**, *31* (2), 342–352. https://doi.org/10.1021/acs.chemmater.8b03272.

(50) Kang, S.; Kim, M.; Min, K. Machine Learning-Aided Discovery of Superionic Solid-State

Electrolyte for Li-Ion Batteries. *arXiv:2202.06763 [cond-mat.mtrl-sci]* **2022**.

(51) Sun, J.; Kang, S.; Kim, J.; Min, K. Accelerated Discovery of Novel Garnet-Type Solid-State Electrolyte Candidates via Machine Learning. *ACS Appl. Mater. Interfaces* **2023**, *15* (4), 5049–5057. https://doi.org/10.1021/acsami.2c15980.

(52) Pereznieto, S.; Jaafreh, R.; Kim, J.; Hamad, K. Solid Electrolytes for Li-Ion Batteries via Machine Learning. *Mater. Lett.* **2023**, *337*, 133926. https://doi.org/10.1016/j.matlet.2023.133926.

(53) Liu, H.; Ma, S.; Wu, J.; Wang, Y.; Wang, X. Recent Advances in Screening Lithium Solid-State Electrolytes Through Machine Learning. *Front. Energy Res.* **2021**, *9* (February), 639741. https://doi.org/10.3389/fenrg.2021.639741.

(54) Fujimura, K.; Seko, A.; Koyama, Y.; Kuwabara, A.; Kishida, I.; Shitara, K.; Fisher, C. A. J.; Moriwake, H.; Tanaka, I. Accelerated Materials Design of Lithium Superionic Conductors Based on First-Principles Calculations and Machine Learning Algorithms. *Adv. Energy Mater.* **2013**, *3* (8), 980–985. https://doi.org/10.1002/aenm.201300060.

(55) Jain, A.; Ong, S. P.; Hautier, G.; Chen, W.; Richards, W. D.; Dacek, S.; Cholia, S.; Gunter, D.; Skinner, D.; Ceder, G.; Persson, K. A. Commentary: The Materials Project: A Materials Genome Approach to Accelerating Materials Innovation. *APL Mater.* **2013**, *1* (1), 011002. https://doi.org/10.1063/1.4812323.

(56) Ong, S. P.; Cholia, S.; Jain, A.; Brafman, M.; Gunter, D.; Ceder, G.; Persson, K. A. The Materials Application Programming Interface (API): A Simple, Flexible and Efficient API for Materials Data Based on REpresentational State Transfer (REST) Principles. *Comput. Mater. Sci.* **2015**, *97*, 209–215. https://doi.org/10.1016/j.commatsci.2014.10.037.

(57) Ong, S. P.; Richards, W. D.; Jain, A.; Hautier, G.; Kocher, M.; Cholia, S.; Gunter, D.;

Chevrier, V. L.; Persson, K. A.; Ceder, G. Python Materials Genomics (Pymatgen): A Robust, Open-Source Python Library for Materials Analysis. *Comput. Mater. Sci.* **2013**, *68*, 314–319. https://doi.org/10.1016/j.commatsci.2012.10.028.

(58) A.Togo. Phonon database (Kyoto university).

(59) Kresse, G.; Furthmüller, J. Efficiency of Ab-Initio Total Energy Calculations for Metals and Semiconductors Using a Plane-Wave Basis Set. *Comput. Mater. Sci.* **1996**, *6* (1), 15–50. https://doi.org/10.1016/0927-0256(96)00008-0.

(60) Kresse, G.; Furthmüller, J. Efficient Iterative Schemes for Ab Initio Total-Energy Calculations Using a Plane-Wave Basis Set. *Phys. Rev. B* **1996**, *54* (16), 11169–11186. https://doi.org/10.1103/PhysRevB.54.11169.

(61) Perdew, J. P.; Burke, K.; Ernzerhof, M. Generalized Gradient Approximation Made Simple. *Phys. Rev. Lett.* **1996**, *77* (18), 3865–3868. https://doi.org/10.1103/PhysRevLett.77.3865.

(62) Belsky, A.; Hellenbrandt, M.; Karen, V. L.; Luksch, P. New Developments in the Inorganic Crystal Structure Database (ICSD): Accessibility in Support of Materials Research and Design. *Acta Crystallogr. Sect. B Struct. Sci.* **2002**, *58* (3), 364–369. https://doi.org/10.1107/S0108768102006948.

(63) Ong, S. P.; Richards, W. D.; Jain, A.; Hautier, G.; Kocher, M.; Cholia, S.; Gunter, D.; Chevrier, V. L.; Persson, K. A.; Ceder, G. Python Materials Genomics (Pymatgen): A Robust, Open-Source Python Library for Materials Analysis. *Comput. Mater. Sci.* **2013**, *68*, 314–319. https://doi.org/10.1016/j.commatsci.2012.10.028.

(64) Togo, A.; Tanaka, I. First Principles Phonon Calculations in Materials Science. *Scr. Mater.* **2015**, *108*, 1–5. https://doi.org/10.1016/j.scriptamat.2015.07.021.

(65) Leo Breiman. Random Forest. *Mach. Learn.* **2001**, *45*, 5–32. https://doi.org/https://doi.org/10.1023/A:1010933404324.

(66) Smith, J. G.; Siegel, D. J. Low-Temperature Paddlewheel Effect in Glassy Solid Electrolytes. *Nat. Commun.* **2020**, *11* (1), 1483. https://doi.org/10.1038/s41467-020-15245-5.

(67) Kraft, M. A.; Culver, S. P.; Calderon, M.; Böcher, F.; Krauskopf, T.; Senyshyn, A.; Dietrich, C.; Zevalkink, A.; Janek, J.; Zeier, W. G. Influence of Lattice Polarizability on the Ionic Conductivity in the Lithium Superionic Argyrodites $Li_6PS_5X$ (X = Cl, Br, I). *J. Am. Chem. Soc.* **2017**, *139* (31), 10909–10918. https://doi.org/10.1021/jacs.7b06327.

(68) de Klerk, N. J. J.; Rosłoń, I.; Wagemaker, M. Diffusion Mechanism of Li Argyrodite Solid Electrolytes for Li-Ion Batteries and Prediction of Optimized Halogen Doping: The Effect of Li Vacancies, Halogens, and Halogen Disorder. *Chem. Mater.* **2016**, *28* (21), 7955–7963. https://doi.org/10.1021/acs.chemmater.6b03630.

(69) Symington, A. R.; Purton, J.; Statham, J.; Molinari, M.; Islam, M. S.; Parker, S. C. Quantifying the Impact of Disorder on Li-Ion and Na-Ion Transport in Perovskite Titanate Solid Electrolytes for Solid-State Batteries. *J. Mater. Chem. A* **2020**, *8* (37), 19603–19611. https://doi.org/10.1039/D0TA05343K.

(70) Di Stefano, D.; Miglio, A.; Robeyns, K.; Filinchuk, Y.; Lechartier, M.; Senyshyn, A.; Ishida, H.; Spannenberger, S.; Prutsch, D.; Lunghammer, S.; Rettenwander, D.; Wilkening, M.; Roling, B.; Kato, Y.; Hautier, G. Superionic Diffusion through Frustrated Energy Landscape. *Chem* **2019**, *5* (9), 2450–2460. https://doi.org/10.1016/j.chempr.2019.07.001.

(71) Sun, Y.; Ouyang, B.; Wang, Y.; Zhang, Y.; Sun, S.; Cai, Z.; Lacivita, V.; Guo, Y.; Ceder,

G. Enhanced Ionic Conductivity and Lack of Paddle-Wheel Effect in Pseudohalogen-Substituted Li Argyrodites. *Matter* **2022**, *5* (12), 4379–4395. https://doi.org/10.1016/j.matt.2022.08.029.

(72) Luo, N.; Hou, Z.; Zheng, C.; Zhang, Y.; Stein, A.; Huang, S.; Truhlar, D. G. Anionic Oxygen Redox in the High-Lithium Material $Li_8SnO_6$. *Chem. Mater.* **2021**, *33* (3), 834–844. https://doi.org/10.1021/acs.chemmater.0c03259.

(73) Kuganathan, N.; Solovjov, A. L.; Vovk, R. V.; Chroneos, A. Defects, Diffusion and Dopants in $Li_8SnO_6$. *Heliyon* **2021**, *7* (7), e07460. https://doi.org/10.1016/j.heliyon.2021.e07460.

(74) Ferraresi, G.; Villevieille, C.; Czekaj, I.; Horisberger, M.; Novák, P.; El Kazzi, M. $SnO_2$ Model Electrode Cycled in Li-Ion Battery Reveals the Formation of $Li_2SnO_3$ and $Li_8SnO_6$ Phases through Conversion Reactions. *ACS Appl. Mater. Interfaces* **2018**, *10* (10), 8712–8720. https://doi.org/10.1021/acsami.7b19481.